\documentclass[3p]{elsarticle}
\usepackage{bm}

\begin{document}

\begin{frontmatter}

\author[SEU]{Rong-Yao Yang\corref{cor1}}
\ead{ryyang@seu.edu.cn}
\author[SEU]{Wei-Zhou Jiang\corref{cor1}}
\ead{wzjiang@seu.edu.cn}
\author[SEU]{Pei-Ying Huo}
\author[APU]{Qi-Lin Zhang}
\affiliation[SEU]{organization={School of Physics, Southeast University}, city={Nanjing}, postcode={211189}, country={China}}
\affiliation[APU]{organization={School of Mathematics-Physics and Finance and School of Materials Science and Engineering, Anhui Polytechnic University}, city={Wuhu}, postcode={241000}, country={China}}
\cortext[cor1]{To whom correspondence should be addressed}

\title{Ultra-efficient mid-infrared energy absorption by water confined in carbon nanotubes}

\begin{abstract}
The energy absorption on nanometer scale is vital for many bio and chemical systems. We report here that a two times amplification in absorption efficiency can be achieved by water molecules confined in carbon nanotubes with small radius, compared with situations in normal bulk water, under irradiations of short mid-infrared pulses. The effect of confinement due to a (6,6) carbon nanotube is found to be very robust, equivalent to that of a 5 $V/nm$ static electric field. These findings are instructive not only for designing high-efficiency nano devices but also for understanding behaviours in biological channels.
\end{abstract}


\end{frontmatter}

\section{Introduction}
The absorptions and emissions of electromagnetic radiations, as one of the most fundamental processes in nature, are crucial for energy transfer and transformation in chemical and biological systems. Photosynthesis is a typical example of energy storage through harvesting light radiation. Mid-infrared (MIR) photon emission was demonstrated from hydrolysis of phosphoanhydride bonds in compounds like adenosine triphosphate (ATP) which is thought to be the direct energy source for life~\cite{2021Li298}. On the other hand, resonant absorptions of artificial radiations were found to have potential applications in opening biological ion channels and modulating neuronal signal~\cite{LiChang-230,LiuQiao-233}. Conformational properties of hydrated macromolecules are very sensitive to electromagnetic irradiations. It was revealed that the DNA unwinding process would experience a 20 times acceleration under a 44 terahertz electric field~\cite{2020Wu296}.
Energy absorption in water systems is of particular interest as it is one of the most important substances on earth to sustain life. The temperature jump achieved by exciting water molecules could induce evolution in chemical reactions~\cite{MaWan-193,2019Cannelli289} and biological dynamics, such as protein unfolding \cite{MunozThompson-279,1998Dyer285,EbbinghausDhar-197}.
Pump-probe experiments based on resonant excitations were widely used to investigate the dynamics in water~\cite{WoutersenEmmerichs-272,2021Yang329}. Though the spectral features in normal bulk water have been well known ~\cite{BertieLan-261,2008Auer330,2009Max353,Mishra2013-11,HuangYang-6,YangHuang-7,MishraBettaque-8}, many properties still remain unexplored for confined water systems interacting with electromagnetic fields on nanometer scale.

Surprising phenomena would occur when a water-involving system is diminished to small size since its dynamical properties could be distinctly different from bulk phase~\cite{2006Dokter388}. For instance, highly toxic viologen compounds, which are stable in bulk water, were found to undergo ultrafast self-degradation in water microdroplets~\cite{2022Gong308}.
Novel square ice phase at room temperature could stably exist when water molecules are constrained between two  hydrophobic graphene sheets~\cite{2015Algara-Siller326}. Particularly, water confined in nanotubes has drawn great attention as its hydrogen-bond network is substantially shielded from fluctuations in the environment, largely different from the case in bulk water~\cite{Hummer2001-18}. In addition, ultrafast conduction of water molecules was found in carbon nanotube (CNT)~\cite{Hummer2001-18}, which makes it possible for applications in desalination of seawater~\cite{2008Corry331}.
It has been demonstrated that water transport through nanotubes is sensitively related with resonant energy absorption under tube or charge vibration and external electromagnetic fields~\cite{2012Rinne347,ZhangJiang-162,2014Kou346,2015Kou343,ZhangYang-163}. Interestingly, the terahertz absorption of nano-confined water was reported to dependent on the pore size of the tube~\cite{2015Sun309}. In this paper, we investigate the energy absorption from MIR radiation by water confined in CNT with different radii and by polarized bulk water under applied static electric field (SEF) using molecular dynamics (MD) simulations. We focus on the excitation of OH stretching vibrations in $H_2O$ molecules. A two times amplification in peak absorption efficiency, compared with the case in bulk water, is found for ordered water chain confined in a (6,6) single-walled CNT (SWCNT). These findings provide essential  information for understanding relative processes in biological channels and developing energy-efficient nano transport devices.

\section{Methods}
\begin{figure}[htbp]
\centering
\includegraphics[height=5.0cm,width=8.0cm]{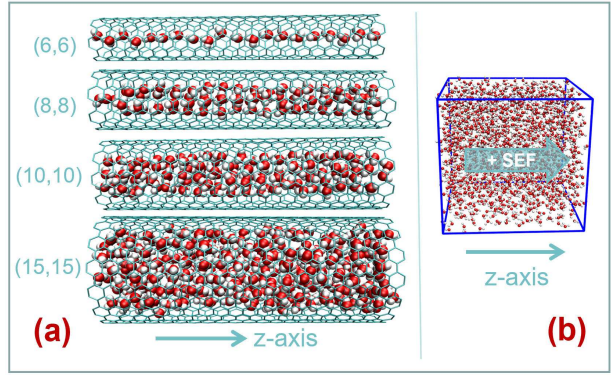}
\caption{(Color online) Snapshots rendered from VMD~\cite{VMD1996} for (a) water molecules in 5.0 $nm$ long SWCNTs with different radii and (b) $3.6\times3.6\times3.6\ nm^3$ supercell for bulk water under polarization of applied SEF along $z$ axis. The red and white balls represent oxygen and hydrogen atoms respectively. The cyan network stands for SWCNTs with carbon atoms at intersections. } \label{F-model}
\end{figure}

We investigate energy absorption of water molecules constrained in SWCNT with different radii under irradiation of MIR pulses. A series of polarized bulk water under various applied SEF constraints were also studied herein for comparison.
The snapshots for 5.0 $nm$ long (6,6), (8,8), (10,10), (15,15) armchair SWCNTs filled with water molecules are shown in Fig.~\ref{F-model}a, containing 20, 69, 127, 362 $H_2O$ molecules respectively. A $3.6\times3.6\times3.6\ nm^3$ cubic supercell with 1560 $H_2O$ molecules, corresponding to a density of 1 $g/cm^3$, is shown in Fig.~\ref{F-model}b. MD simulations were conducted under periodic boundary conditions using NAMD package~\cite{NAMD2005,NAMD2020} with 0.2 $fs$ time step, charmm27 force field~\cite{Charmm1998} and flexible TIP3P water model~\cite{TIP3P1983}. We mention that TIP3P model has captured the main physics about exciting OH stretching vibrations in water molecules and gives the right frequency domain of peak absorption, which justifies the usage of this model to illuminate the enhancement of the HO vibrational excitation under confinement, though it can not reproduce the experimental line shape due to absence of coupling between intra and inter molecular vibrations. A 1.2 $nm$ cutoff was applied for the van der Waals interactions and electrostatic interactions which were handled by the particle mesh Ewald algorithm~\cite{PME1993}. To mimic polarized bulk water, a SEF of 1 or 5 $V/nm$ was exerted on the supercell in Fig.~\ref{F-model}b throughout the simulation, while the case without SEF represents normal bulk water.
The simulations in general comprise two steps: (1) a 1 $ns$ long canonical equilibrium simulation at 300 $K$ was performed under Langevin thermostat, (2) a subsequent 5 $ps$ non-equilibrium simulation was carried out with temperature control withdrawn to evaluate the energy absorption during the irradiation of MIR pulses.
The electric field according to the pulses polarized along $z$ axis is written as ${\bf E} = [0,\ 0,\ E_0]$ where $E_0=A\cdot cos(2\pi f t)\cdot e^{(t-t_c)^2/2\sigma^2}$ with $f$ being the frequency of the incident laser pulse, $t$ being the time in the simulation. The relationship between frequency and  wavenumber is $f=c\tilde{\nu}$ where $c$ is the speed of light in vacuum.
Unless otherwise mentioned, the parameters were set to $A=0.5\ V/nm$, $\sigma=1/2.355\ ps$ (denoting $1\ ps$ full width at half maximum (FWHM)) and $t_c=2.5\ ps$. The insignificant Lorentz force due to magnetic field in laser pulses is neglected here. In this paper, we focus on the MIR absorption based on resonant excitation of water's OH stretching vibrations.
The energy absorption (E$_{absorb}$) per $H_2O$ molecule is defined as the total energy difference between final state ($t=5\ ps$) and initial state ($t=0\ ps$) in the non-equilibrium simulation divided by the number of $H_2O$ molecules in the system.

\section{Results and discussion}
\begin{figure}[htbp]
\centering
\includegraphics[height=6.0cm,width=8.0cm]{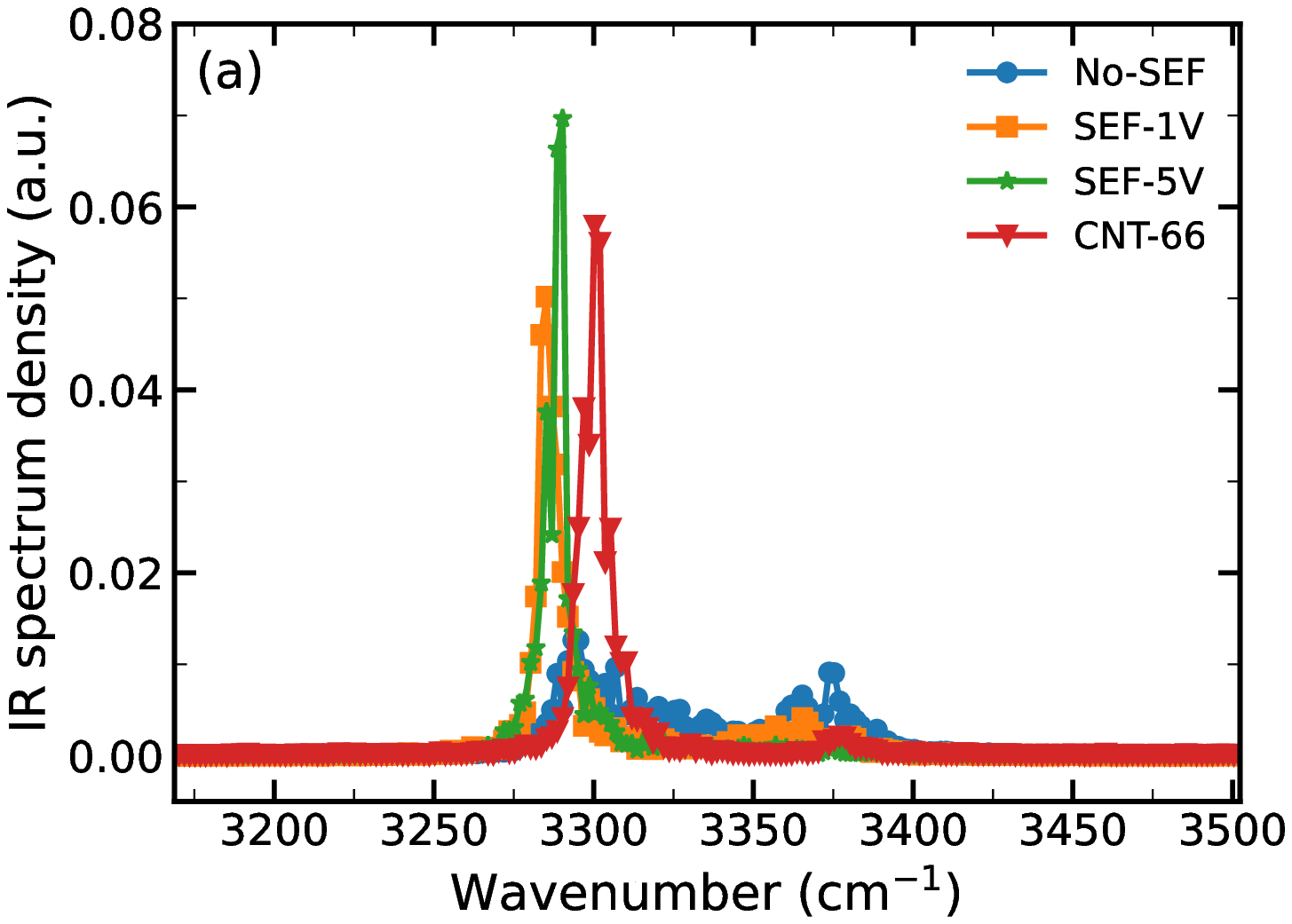}
\includegraphics[height=6.0cm,width=8.0cm]{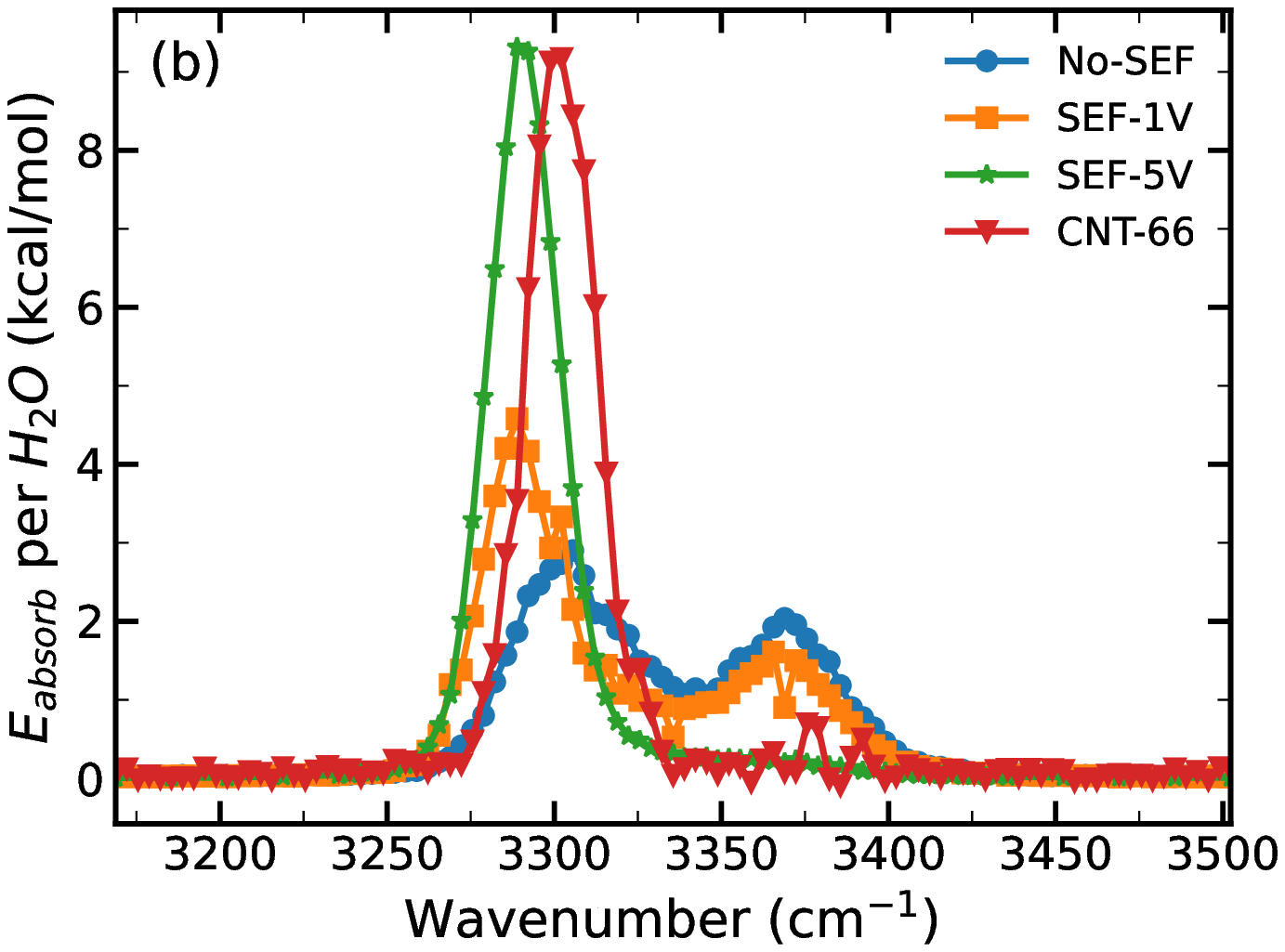}
\caption{(Color online) (a) IR spectrum density along $z$ axis for normal bulk water (No-SEF), polarized bulk water under 1 or 5 $V/nm$ SEF constraint (SEF-1V or SEF-5V), and water chain confined in a (6,6) SWCNT (CNT-66). (b) Corresponding energy absorption per $H_2O$ molecule achieved under irradiation of pulses with different wavenumbers. The abbreviations in the legends are kept the same in the following.} \label{F-IRSpectra-E-rise}
\end{figure}

Properties of MIR pulses' absorption can usually be inferred from the infrared (IR) spectrum based on dipole autocorrelation function. Displayed in Fig.~\ref{F-IRSpectra-E-rise}a are the IR spectrum densities along $z$ axis calculated from microcanonical trajectories  in frequency domain of OH stretching vibrations for normal bulk water, polarized bulk water under 1 or 5 $V/nm$ SEF constraint, and water chain confined in a (6,6) SWCNT. Prominent peaks can be spotted around 3300 $cm^{-1}$ for polarized bulk water and water chain in (6,6) SWCNT whose peak values are more than four time that of the normal bulk water. These spectrum structures predict that resonant absorptions shall occur under irradiation of pulses near 3300 $cm^{-1}$ and water systems constrained by high intensity SEF or CNT would exhibit considerable enhancement, compared with norm bulk water.

These spectral features are demonstrated by the non-equilibrium MD simulations with its results of energy absorption shown in Fig.~\ref{F-IRSpectra-E-rise}b. The frequency locations of peak absorptions are consistent with Fig.~\ref{F-IRSpectra-E-rise}a, while the line width of the absorption is broadened due to the dynamical evolving of water morphoses during the irradiation of pulses. The peak value of energy absorption per $H_2O$ molecule is 2.9 $kcal/mol$ at 3300 $cm^{-1}$ for normal bulk water. This peak value is much less than 9.3 $kcal/mol$
for polarized bulk water under 5 $V/nm$ SEF and 9.2 $kcal/mol$
for water chain in (6,6) SWCNT. The peak value for polarized water under 1 $V/nm$ SEF is 4.6 $kcal/mol$ which is substantially smaller than the prediction from IR spectrum density. In the aspect of peck absorption efficiency, the effect due to confinement of (6,6) SWCNT is approximately equivalent to 5 $V/nm$ SEF constraint. However, an intense external electric field is less practicable in water systems since water molecules are able to undergo dissociation when the SEF reaches 3.5 $V/nm$~\cite{Saitta2012-2}.
The significant enhancement of energy absorption in strongly confined water systems is very interesting as it illustrates a highly efficient pathway in nano or bio systems.

\begin{figure}[htbp]
\centering
\includegraphics[height=6.0cm,width=8.0cm]{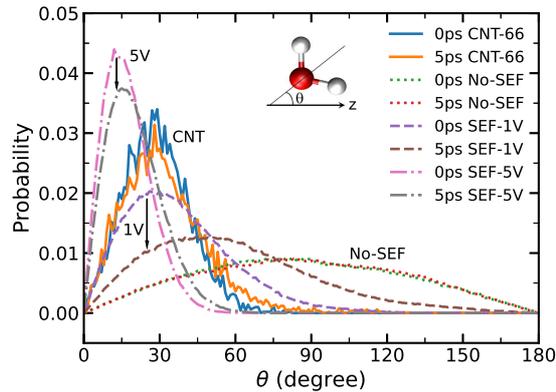}
\caption{(Color online) Probability distributions of the angle between water's axis of symmetry and $z$ axis for the water in (6,6) SWCNT, polarized bulk water under 1 or 5 $V/nm$ SEF constraint and  normal bulk water. The labels ``0ps'' and ``5ps'' represent results counted from initial state at 300 $K$ ($t=0\ ps$) and from heated final state ($t=5\ ps$) after irradiation of a 3300 $cm^{-1}$ pulse in the non-equilibrium simulations, respectively. } \label{F-ang}
\end{figure}

The resonant absorption coefficient is determined by the transition probability of a molecule from ground state $|i\rangle$ to excited state $|f\rangle$, which is proportional to $|\langle i|\bm{E\cdot\mu}|f\rangle|^2$ according to Fermi's Golden Rule~\cite{VitaleDziedzic-207}, where $\bm{\mu}$ is the operator of dipole moment and $\bm{E\cdot\mu}=E_0\mu\cos\theta$ with $\theta$ being the angle between transition dipole moment and $z$ axis in the cases concerned here. Therefor, the efficiency of energy absorption is positively correlated with the ensemble average of $\cos^2\theta$. For the excitation of symmetric OH stretching vibration, the angular distribution of dipole moment of $H_2O$ molecules along $z$ axis could provides a preliminary perception on the relative magnitude of the energy absorptions.
Shown in Fig.~\ref{F-ang} are the initial (t=0 $ps$) and final  (t=5 $ps$) angular probability distributions for different water systems. Compared with the case in normal bulk water, a characteristic of small-angle preference in the distribution (i.e., tend to parallel with $z$ axis) is noticeable for the water in (6,6) SWCNT or under strong SEF polarization, which indicates more efficient absorption due to the confinements. The initial and final angular distributions in (6,6) SWCNT almost remain unchanged during the irradiation of incident pulses, while the small-angle population decreases considerably for the bulk water under 1 $V/nm$ SEF and reduces moderately for the case with 5 $V/nm$.
This orientational change during the dynamic processes of pulse absorption is accountable for the discrepancy between the IR spectrum prediction in Fig.~\ref{F-IRSpectra-E-rise}a and the energy absorbed from pulses in Fig.~\ref{F-IRSpectra-E-rise}b. The robust confinement in (6,6) CNT insures the super-high absorption efficiency and a less rigorous constraint by 1 $V/nm$ SEF fails to sustain the process of fierce energy absorption. We mention that the capability of CNT to support concrete confinement is foreseeable since an internal pressure can reach as high as 40 gigapascals inside a CNT~\cite{2006Sun325}. However, it is quite startling that the ordered water structure in CNT can withstand violent energy absorption under  irradiation of intense MIR pulses. Since ordered water network facilitates high-efficiency electron transfer and energy redistribution~\cite{2016Perakis394,Wang2021-146}, this strong structural maintainability shall have potential applications in catalysis and energy science.


\begin{figure}[htbp]
\centering
\includegraphics[height=6.0cm,width=8.0cm]{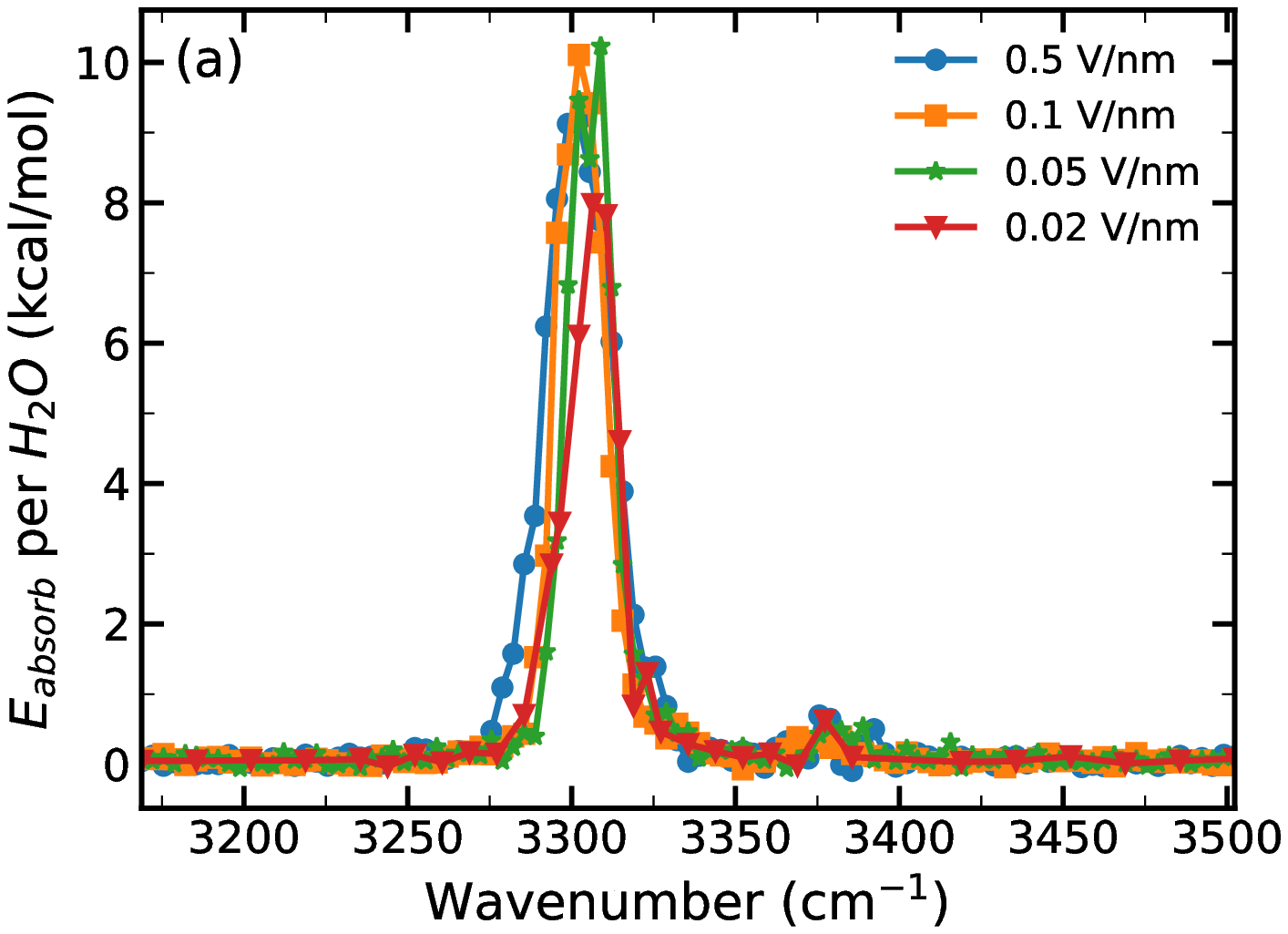}
\includegraphics[height=6.0cm,width=8.0cm]{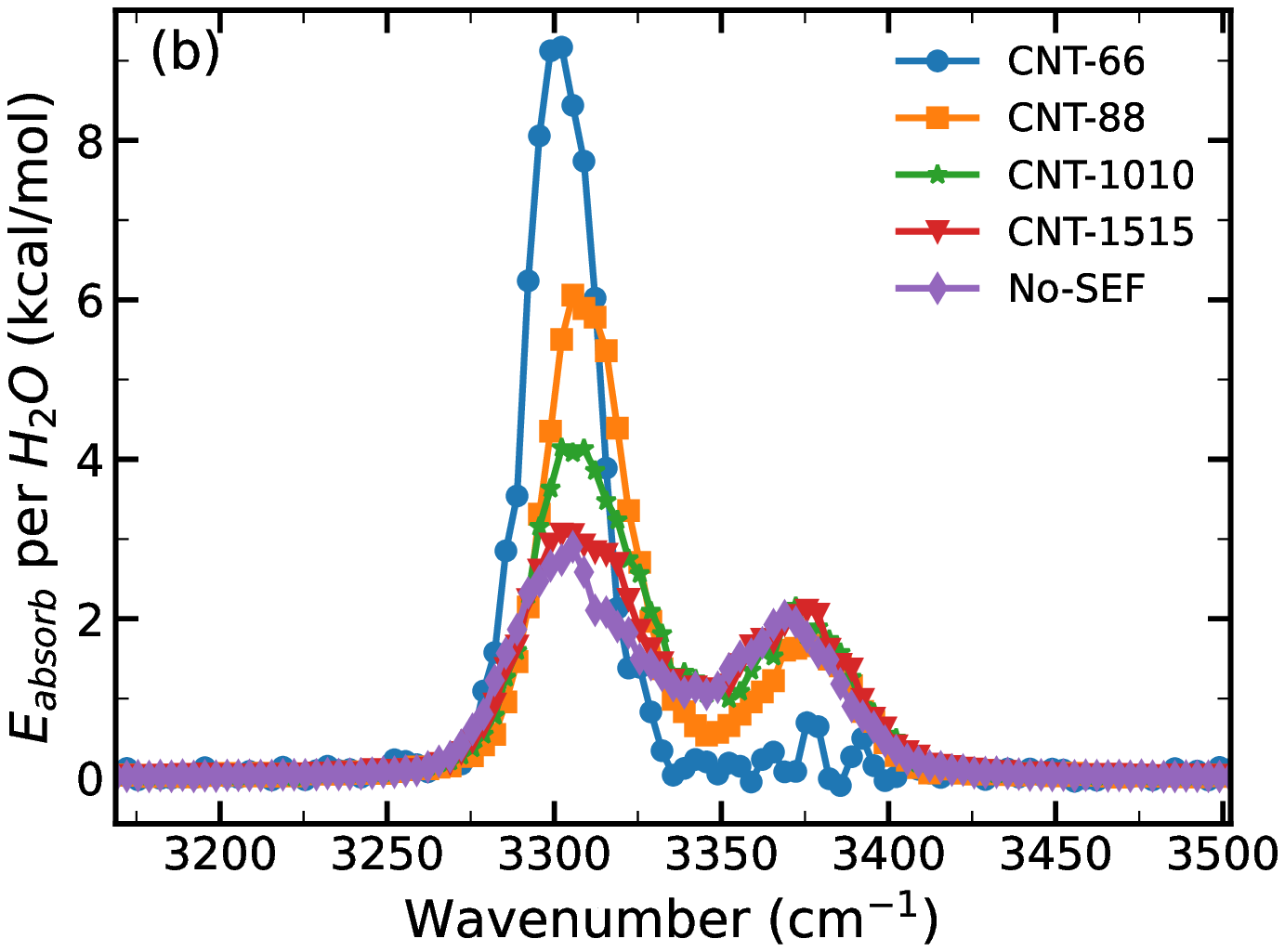}
\caption{(Color online) Energy absorption per $H_2O$ molecule achieved by (a) water in (6,6) SWCNT with the maximum intensities of incident pulses being  $A=0.5,\ 0.1,\ 0.05,\ 0.02\ V/nm$ and (b) water in (6,6), (8,8), (10,10), (15,15) SWCNTs under irradiation of $A=0.5 V/nm$ pulses.} \label{F-EF-radii}
\end{figure}

The effect of maximum intensity of pulse is clarified in Fig.~\ref{F-EF-radii}a with results collected from non-equilibrium simulations using $A=0.5,\ 0.1,\ 0.05,\ 0.02\ V/nm$ and the corresponding FWHM$=1,\ 25,\ 100,\ 625\ ps$ respectively (to keep the total energy of the incident pulses unchanged). The amounts of energy absorbed per $H_2O$ molecule are very similar with each other for these cases, which means that the effect of confinement in (6,6) CNT can sustain for a long time. Increasing the radius of the CNT will release the radial confinement and water morphosis will gradually resemble that in bulk water. As can be seen from Fig.~\ref{F-EF-radii}b, the peak efficiency of energy absorption decreases apparently with increasing the CNT radius. For water in a (15,15) CNT, the profile of energy absorption is comparable to that of normal bulk water.

\section{Conclusion}

The resonant energy absorption from MIR pulses by water systems confined in CNTs or constrained by intense SEF are investigated using MD simulations. Compared with the case in normal bulk water, more than two times amplification in absorption efficiency is achieved by water chain in a (6,6) CNT and polarized bulk water under 5 $V/nm$ SEF. These strong confinements reorder water molecules to morphoses with a highly-biased alignment, which offers a chance to increase the transition probability of exciting the symmetric OH stretching vibration when the direction of the incident pulse is set to be parallel with the preferential orientation of water molecules. Meanwhile, the confinement of a (6,6) CNT is robust enough to maintain the ordered water structure even after fierce energy absorption owing to intense MIR irradiation. These findings emphasis the importance of confinements in water systems and are instructive not only for designing high-efficiency nano devices but also for comprehending relevant phenomena in biological channels.

\section*{acknowledgement}
R.-Y. Yang sincerely thank Dr. Hua Chen for useful discussions. We appreciate the financial support from the National Natural Science Foundation of China under Grant No. 11775049 and the China Postdoctoral Science Foundation under Grant No. 2021M690627.

\bibliographystyle{elsarticle-num}
\bibliography{ref-cntHeat}

\end{document}